\documentclass[journal]{IEEEtran}
\usepackage{}
\usepackage{amssymb}
\usepackage{graphicx,cite,epsfig,amssymb,amsmath}
\usepackage{epstopdf}

\usepackage{pifont}
\usepackage{stmaryrd}
\usepackage{bbding}
\usepackage{algorithm}
\usepackage{algorithmic}
\usepackage{mathrsfs}
\usepackage{latexsym}
\usepackage{indentfirst}
\usepackage{fmtcount}
\usepackage{cite}
\usepackage{float}
\usepackage{morefloats}

\usepackage{caption}
\usepackage{subcaption}

\usepackage{color}
\usepackage{mathtools}
\usepackage{xfrac}

\usepackage{array,colortbl,multirow,multicol,booktabs,ctable,bigstrut}

\usepackage[
top    = 0.75in,
bottom = 0.75in,
left   = 0.75in,
right  = 0.75in]{geometry}
\usepackage{fancyhdr}

\begin{document}

\title{Reinforcement Learning for Nested Polar Code Construction}

\author{
    \IEEEauthorblockN{Lingchen~Huang\IEEEauthorrefmark{1}, Huazi~Zhang\IEEEauthorrefmark{1}, Rong~Li\IEEEauthorrefmark{1}, Yiqun~Ge\IEEEauthorrefmark{2}, Jun~Wang\IEEEauthorrefmark{1}}\\
    \IEEEauthorblockA{\IEEEauthorrefmark{1}Hangzhou Research Center, Huawei Technologies, Hangzhou, China}\\
    \IEEEauthorblockA{\IEEEauthorrefmark{2}Ottawa Research Center, Huawei Technologies, Ottawa, Canada}\\
    Email: \{huanglingchen,zhanghuazi,lirongone.li,yiqun.ge,justin.wangjun\}@huawei.com
}
\maketitle

\begin{abstract}
In this paper, we model nested polar code construction as a Markov decision process (MDP), and tackle it with advanced reinforcement learning (RL) techniques. First, an MDP environment with state, action, and reward is defined in the context of polar coding. Specifically, a state represents the construction of an $(N,K)$ polar code, an action specifies its reduction to an $(N,K-1)$ subcode, and reward is the decoding performance. A neural network architecture consisting of both policy and value networks is proposed to generate actions based on the observed states, aiming at maximizing the overall rewards. A loss function is defined to trade off between exploitation and exploration. To further improve learning efficiency and quality, an ``integrated learning'' paradigm is proposed. It first employs a genetic algorithm to generate a population of (sub-)optimal polar codes for each $(N,K)$, and then uses them as prior knowledge to refine the policy in RL. Such a paradigm is shown to accelerate the training process, and converge at better performances. Simulation results show that the proposed learning-based polar constructions achieve comparable, or even better, performances than the state of the art under successive cancellation list (SCL) decoders. Last but not least, this is achieved without exploiting any expert knowledge from polar coding theory in the learning algorithms.
\end{abstract}


\IEEEpeerreviewmaketitle

\section{Introduction}\label{section:intro}
In communication systems, the capacity of an AWGN channel is defined in theory~\cite{Shannon}.
Classic code construction methods are built upon coding theory, in which code performance can be theoretically modeled in terms of various types of code properties, e.g. minimum distance, decoding threshold, reliability, etc.
However, it seems insufficient for us to rely on only these classic coding theory metrics in facing of such practical concerns and application-specific requirements as realistic channel types, decoding latency and complexity and so on.

Recently, artificial intelligence (AI) techniques have been applied to physical layer design.
AI techniques can be a tool to design or optimize error correction codes~\cite{AI:AI_coding}, while leaving their legacy encoding and decoding architectures and implementations unchanged.
Within a ``constructor-evaluator'' framework~\cite{AI:AI_coding}, AI algorithms such as policy gradient, genetic algorithm, and actor critic, are capable of constructing linear block codes and polar codes with as good performances as the state of the art.
In~\cite{AI:RL_LDPC}, RL and Monte Carlo tree search (MCTS) are combined to guide edge growth in LDPC code construction.
In~\cite{AI:GeneAlg_Polar,AI:GeneAlg_LDPC}, genetic algorithms are used to design polar codes and LDPC codes. The main difference from~\cite{AI:AI_coding} is that coding expert knowledge is utilized during the initialization to speed up the learning process.

In this paper, our motivation is to investigate the feasibility of using AI technologies to explore the design space for wireless systems. Channel code, especially polar code, is a good example for this endeavor.
We propose novel RL algorithms for designing nested polar codes~\cite{Polar:nested}.
Because nested polar code construction (sequential information sub-channel selection) is inherently modeled as a Markov decision process (MDP), and RL algorithms can be applied to approach the optimum.
To improve training efficiency and code performance, we propose an integrated learning paradigm and various parameter optimization techniques.

The structure of this paper is as following.
Section~\ref{section:motivation} introduces the preliminaries about polar code construction and nested polar codes. Section~\ref{section:learning} models the nested polar code construction as an MDP task, and solves it with several advanced reinforcement learning algorithms. The integrated learning paradigm is introduced in Section~\ref{section:int_learning}.
All proposed algorithms are evaluated in Section~\ref{section:evaluation} in terms of sample efficiency and code performance.
Conclusions are given in section~\ref{section:conclusion}.

\section{Preliminaries}\label{section:motivation}
\subsection{Polar code construction}\label{section:motivation:polar}
Polar codes~\cite{Polar:Arikan} are the first class of capacity-achieving codes (under successive cancellation (SC) decoding).
For polar codes, physical channels are synthesized to polarized subchannels, with the most reliable ones selected to carry information bits.
As a result, an $(N,K)$ polar code is defined by the $K$ most reliable subchannel indices, namely information set $\cal I$.
The remaining $(N-K)$ subchannel indices are defined as frozen set $\cal F$.
As code length $N$ increases, subchannels polarize to either purely noiseless or completely noisy, where the fraction of noiseless subchannels approaches channel capacity~\cite{Polar:Arikan}.
For binary erasure channel (BEC), subchannel reliability can be efficiently calculated by Bhattacharyya parameter.
For general binary-input memoryless channels, density evolution (DE) was applied to estimate subchannel reliability~\cite{Polar:DE1_Mori,Polar:DE2_Mori}, and improved in~\cite{Polar:DE3_Tal} and analyzed in~\cite{Polar:DE4_Pedarsani} in terms of complexity.
For AWGN channels, Gaussian approximation to density evolution (DE/GA) was proposed~\cite{Polar:GA_Trifonov} to further reduce complexity with negligible performance loss.

To improve the performance of polar codes at finite length, enhanced decoding algorithms are proposed~\cite{Polar:SCL,Polar:SCS}.
Among them, SC list (SCL) decoding achieves the best tradeoff among decoding latency, complexity and performance.
However, to our best knowledge, for polar codes with SCL-based decoders, theoretically optimal code construction is still an open problem.
Existing constructions either directly adopt DE/GA, which are designed for SC rather than SCL, or apply genetic algorithms for SCL decodings~\cite{AI:AI_coding,AI:GeneAlg_Polar}.

\subsection{Nested polar codes}\label{section:motivation:nested_polar}
In practical communication systems where code rate and length adaption is required, efficient code description is mandatory.
For example, 5G enhanced mobile broadband (eMBB)~\cite{Polar:212} supports thousands of polar codes with different $(N,K)$ combinations.
It is impossible to store all code configurations separately, due to large overhead.
It is much more convenient for description and implementation to impose a nested property~\cite{Polar:nested}, so that all polar codes of the same mother code length can be derived from a single nested sequence.
Specifically, denote ${\cal F}_{N,K}$ as the frozen set of an $(N,K)$ polar code.
${\cal F}_{N,N-1}, {\cal F}_{N,N-2}, \cdots, {\cal F}_{N,0}$ can be constructed sequentially, on condition that ${\cal F}_{N,N-1} \subset {\cal F}_{N,N-2} \subset \cdots \subset {\cal F}_{N,0}$.
As seen, a single nested sequence can be obtained as $\{{\cal F}_{N,N-1}, \textit{setdiff}( {\cal F}_{N,N-2}, {\cal F}_{N,N-1} ), \cdots, \textit{setdiff}( {\cal F}_{N,0}, {\cal F}_{N,1})\}$.
\begin{figure}[h]
\centering
    \includegraphics[width = 0.45\textwidth]{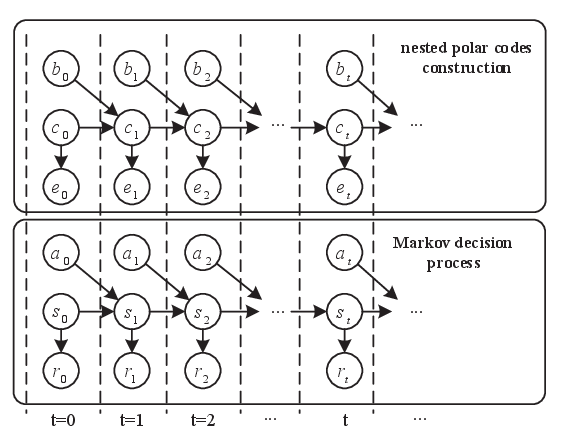}
    \caption{Graphical model of nested polar code construction and Markov Decision Process. For nested polar code construction, $c$ denotes a polar code, $b$ denotes a subchannel, $e$ denotes error correction performance with respect to (w.r.t.) $c$. For MDP, $s$ denotes a state, $a$ denotes an action, and $r$ denotes a reward value w.r.t. the state $s$.}
    \label{fig:graphical_models}
\end{figure}

Nested polar codes are adopted by 5G in the form of a reliability sequence of length $N_{\max}=1024$~\cite{Polar:212}.
To construct an $(N,K)$ polar code from the length-$N_{\max}$ nested sequence ($N=2^n \leq N_{\max}$),
\begin{enumerate}
  \item First, a sequence $Seq_{N}$ of length $N$ is  extracted from the length-$N_{\max}$ sequence (by taking all indices $\{i:i\in Seq_{N_{\max}},i<N\}$ while keeping the ordering).
  \item Second, the last $K$ entries of $Seq_{N}$ are selected as the information set.
\end{enumerate}

\section{Reinforcement Learning for nested polar code construction}\label{section:learning}
In this section, we show that nested polar code construction is actually a Markov decision process (MDP) that can be tackled by reinforcement learning.
We further discuss some applicable learning algorithms.
\subsection{Constructing nested polar code with MDP}\label{section:learning:MDP}
Nested polar code construction can be modeled as an MDP for the following reasons:
\begin{itemize}
  \item According to Markov property of nested polar code construction in Fig.~\ref{fig:graphical_models}, the construction of $(N,N-K-1)$ polar code $c_{K+1}$ and its performance $e_{K+1}$ depend only on that of $(N,N-K)$ polar code $c_{K}$ and a subchannel selection $b_K$;
  \item The goal is to optimize all $(N,K)$ polar codes for $K=1,2,\cdots,N-1$ through maximizing an overall performance metric $\sum_K{e_K}$.
\end{itemize}

To explicitly map the nested polar code construction into an MDP task, we define a base environment $\left( \mathcal {S}, \mathcal {A}, \mathcal {R} \right)$:
\begin{itemize}
  \item A \textbf{state} is denoted by a length-$N$ binary vector $s_K^N \in \mathcal {S}=\{0,1\}^N$, whose support set is ${\cal F}_{N,N-K}$. The initial state $s_0^N$ is an all-zero vector corresponding to empty set (${\cal F}_{N,N}$).
  \item An \textbf{action} is denoted by an integer $a_K^N \in \mathcal {A}=\{0,1,...,N-1\}$, such that $a_K^N \notin {\cal F}_{N,N-K}$ and $a_K^N \cup {\cal F}_{N,N-K} = {\cal F}_{N,N-K-1}$.
  \item The \textbf{reward} value of state $s_K^N$ is $r_K^N \in \mathcal {R}$, representing the performance of the polar code defined by ${\cal F}_{N,N-K}$.
\end{itemize}
The state transfer process is deterministic, i.e., given $s_K^N$ and $a_K^N$, $s_{K+1}^N$ can be determined.
The maximum length of an episode is $N$.
A trajectory of the base environment, $(s_0^N$, $a_0^N$, $s_1^N$, $a_1^N$, $\cdots$, $s_{N-1}^N$, $a_{N-1}^N$, $s_{N}^N)$, corresponds to the nested polar code construction (ordered sequence) $\{a_0^N$, $a_1^N$, $\cdots$, $a_{N-1}^N\}$.
In the following, the superscripts of $s_K^N$, $a_K^N$ and $r_K^N$ are omitted with some abuse of notation.

Following the ``constructor-evaluator'' framework \cite{AI:AI_coding}, we propose to directly evaluate the rewards through decoding performance. Monte-Carlo (MC) simulations are conducted to output a block error rate (BLER) performance for each code construction.
The evaluator implements SCL decoding algorithms, which generate a list of $L$ codewords.
We name two types of SCL decoders based on final output selection:
\begin{itemize}
  \item SCL-PM: select the first codeword, i.e, the most likely one with the smallest PM;
  \item SCL-Genie: select the correct codeword, as long as it is among the $L$ surviving ones.
\end{itemize}
Sufficient decoding error events are counted to obtain an accurate BLER estimation.
Then, the reward value is defined as $r\triangleq-\log_{10}BLER$.
\footnote{For SCL-Genie decoder, when $K \leq \log_{2}L$, the codeword would always be decoded correctly. For such cases, the reward value is set to $0$.}

\subsection{Reinforcement learning}\label{section:learning:RL_algs}
Nested code construction is actually to search an optimal sequence in a large solution space. RL would help approach the optimum, dragged by a reward. The devised reward, through one real value metric, should represent the performances of all component codes.

Strictly speaking, an RL agent interacts with the MDP environment over discrete timesteps. At each timestep $t$,
the agent observes a state $s_t$, chooses an action $a_t$ according to its policy $\pi(a_t|s_t)$ and obtains a reward $r_t$ from the environment.
The goal of this agent is to optimize its policy in order to maximize the discounted return $R_t=\sum_{i=0}^{\infty} \gamma^{i} r_{t+i}$ at each timestep.
Here discount factor $\gamma \in [0,1)$ is introduced to trade off the contribution of immediate and long term rewards to return value.

For the nested polar code construction task, the state space is $2^{N}$, the action space is $N$ and the solution space, i.e. trajectory space, is $N!$.
Concerning the large solution space, it is necessary to have \emph{sample efficient} RL algorithms.
\emph{Sample efficiency} is defined by number of samples used to solve the MDP task, where an MDP sample is a state-action-reward $(s,a,r)$ tuple.
In literature, sample efficient RL algorithms include advantage actor critic (A2C), proximal policy optimization (PPO)~\cite{RL:PPO} and actor critic using Kronecker-factored trust region (ACKTR) \cite{RL:ACKTR}, etc.

We apply PPO~\cite{RL:PPO} as it is by far the most advanced model-free algorithm.
The PPO is an extension of A2C, where a Kullback-Leibler (KL) divergence constraint is imposed between the updated policy and the old policy, i.e. a trust region constraint~\cite{RL:TRPO}.

For the PPO, the policy loss function is defined,
\begin{equation}\label{Equ.actor}
  Loss_{A} = \hat{A}(s_t,a_t) \cdot min\left(r_t(\theta), clip(r_t(\theta), 1-\epsilon, 1+\epsilon)\right),
\end{equation}
where $\hat{A}(s_t,a_t)=R(s_t,a_t)-V(s_t)$ is the estimate of advantage function for taking action $a_t$ at state $s_t$;
$r_t(\theta)=\frac{\pi_{\theta}(a_t|s_t)}{\pi_{\theta_{old}}(a_t|s_t)}$ is the probability ratio between the updated policy $\pi_{\theta}$ and the old policy $\pi_{\theta_{old}}$ for taking action $a_t$ at state $s_t$;
$\pi_{\theta}(a_t|s_t)$ is the policy function parameterized by $\theta$;
$\epsilon$ is a clipping ratio to constrain the probability ratio $r_t(\theta)$.

The value loss function is defined,
\begin{equation}\label{Equ.critic}
  Loss_{C} = \left( \hat{A}(s_t,a_t) \right)^2.
\end{equation}

For the advantage estimation $\hat{A}(s_t,a_t)$, a general advantage estimation method (GAE)~\cite{RL:GAE} implements an exponential average among advantage estimations of different steps to trade off between the estimation bias and variance,
\begin{equation}\label{Equ.GAE}
\begin{aligned}
  \hat{A}^{GAE}(s_t,a_t)& = (1-\lambda) \sum_{i=1}\left( \lambda^{i-1} \hat{A}^{i}(s_t,a_t) \right), \\
  \hat{A}^{i}(s_t,a_t)& = \hat{R}^{i}(s_t,a_t) - V(s_t), \\
  \hat{R}^{i}(s_t,a_t)& = \sum_{j=0}^{i-1}\left( \gamma^{j} r_{t+j} \right) + \gamma^{i} V^{\pi}(s_{t+i}),
\end{aligned}
\end{equation}
where $\lambda$ is the exponential moving average parameter.

A policy function entropy regularization, defined in \eqref{Equ.entropy}, can be considered in policy loss function to trade off between exploration and exploitation.
\begin{equation}\label{Equ.entropy}
  H_{A}(s) = -\sum_{a}\pi(a|s)\log \pi(a|s).
\end{equation}

\section{Integrated learning for nested polar code construction}\label{section:int_learning}
In the section, we propose an integrated learning method for nested polar code construction to improve the sample efficiency and code performance.

For reinforcement learning algorithms, policy function is initialized to explore all possible MDP trajectories with equal probability.
However, for most trajectories in the trajectory space, the accumulated rewards are far worse than optimal one(s).
Given prior knowledge about the distributions of actions with large rewards, the policy function can be pretrained to bias the exploration towards trajectories with larger accumulated rewards.
Depending on the prior knowledge, this pretraining can significantly accelerate the learning process~\cite{RL:Pretrain1,RL:Pretrain2,RL:Pretrain3}.

In the context of polar code construction, we may rely on sub-optimal expert knowledge (e.g., DE/GA constructions) for pretraining, where direct state-action $(s,a)$ pairs (demonstrations) are available.
However, genetic algorithm is the best choice to generate a large population of (sub-)optimal code constructions, corresponding to the distribution of states with large rewards. As the genetic algorithm converges, its population already contains code constructions with the best performances. Moreover, the genetic algorithm in \cite{AI:AI_coding} does not require any expert knowledge, which means the proposed method also learns everything from scratch.

An integrated learning is proposed in Alg.~\ref{alg:integrated_learning}.
Firstly, the polar code constructions are generated by genetic algorithm. 
These constructions are used to produce pretraining examples. 
Policy function is then pretrained in supervised learning manner.
Nested polar code constructions are learned through reinforcement learning, as in section~\ref{section:learning:RL_algs}, with the pretrained policy function.

\begin{algorithm}
\begin{algorithmic}[1]
\STATE $polar\_codes\_constr$ = genetic\_algorithm()
\STATE $(state, action)$ = \\example\_generation($polar\_codes\_constr$)
\STATE $policy\_function\_pretrained$ = \\pretrain($policy\_function, state, action$)
\STATE $nested\_polar\_codes\_constr$ = \\reinforcement\_learning($policy\_function\_pretrained$)
\end{algorithmic}
\caption{Integrated learning algorithm for nested polar code construction}
\label{alg:integrated_learning}
\end{algorithm}

\subsection{Multi-stage genetic algorithm}\label{section:int_learning:GA}
For each $(N,K)$ pair, we apply the genetic algorithm in~\cite{AI:AI_coding} to generate a population of (sub-)optimal polar codes.
In the original version \cite{AI:AI_coding}, BLER performance is evaluated at a fixed SNR, where existing code constructions achieve $BLER\approx(10^{-2}\sim10^{-3})$.
This setting causes a ``slow start'' problem during the beginning phase when most code constructions result in $BLER=1$, which means equally bad.
This would confuse the genetic algorithm as it could not distinguish good code constructions from bad ones, and loses the direction of evolution.
As a result, the genetic algorithm either stucks at this phase, or converges very slowly.

\begin{figure}[h]
\centering
    \includegraphics[width = 0.45\textwidth]{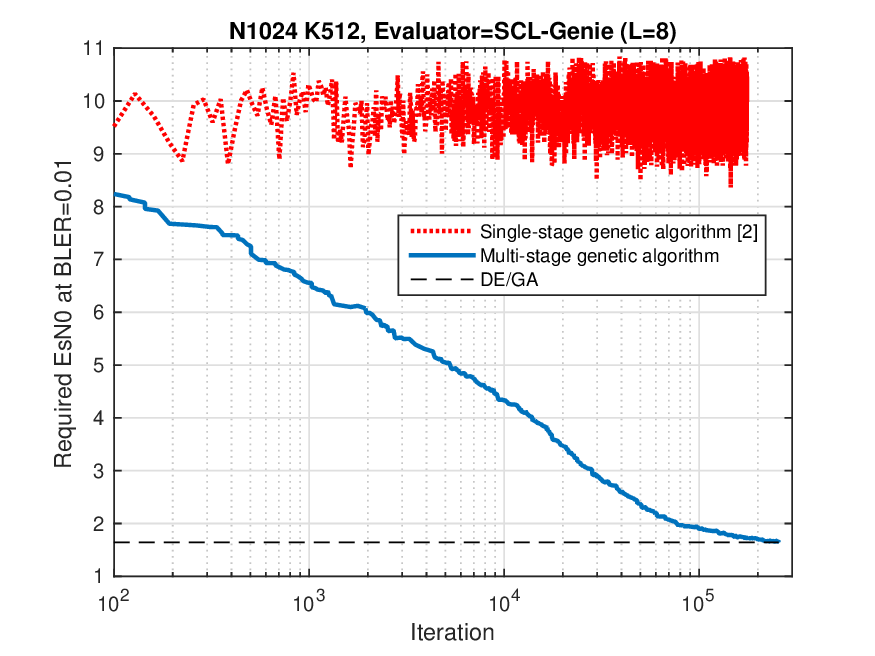}
    \caption{The convergence of multi-stage genetic algorithm for learning a $(N=1024, K=512)$ code.}
    \label{fig:multistage}
\end{figure}

In this work, we propose a novel multi-stage genetic algorithm based on \cite{AI:AI_coding} to improve learning efficiency.
The idea is simple, i.e., to adaptively set the evaluating SNR such that the BLER performances of different code constructions can be differentiated.
Specifically, the evaluating SNR decreases in a multi-stage manner, by tracking the working SNR (at $BLER\approx10^{-2}\sim10^{-3}$) of the best code construction in the population.
The algorithm is detailed in Alg.~\ref{alg:genetic-multistage}. It enables fast convergence especially for longer codes. As shown in Fig.~\ref{fig:multistage}, the construction of a $(N=1024, K=512)$ code is learned to perform well under SCL-Genie with list size $L=8$. In contrast, a single-stage genetic algorithm fails to converge within a reasonable time period.

\begin{algorithm}
\begin{algorithmic}[1]
\STATE {\bf function} genetic\_algorithm()
    \STATE Parameters: population size $M=N$, sample focus $\alpha=0.1$, mutation rate $\beta=0.7$, SNR step $SNR_{step}=0.5$;
    \STATE Initialize population by randomly selected information subchannels: ${\cal I}_1, {\cal I}_2, \cdots, {\cal I}_M$;
    \STATE Sort population by decoding performance: ascending BLER at $SNR_{eval}=SNR_{\max}$;
    \WHILE 1
        \IF {The best code construction ${\cal I}^*$ has $BLER^*<10^{-3}$}
            \STATE Set $SNR_{eval}=SNR_{eval}-SNR_{step}$
            \STATE Re-sort population by decoding performance: ascending BLER at $SNR_{eval}$;
        \ENDIF
        \STATE Select parents ${\cal I}_{p1}, {\cal I}_{p2}$ from population according to fitness, e.g., the $i$-th one is selected with probability $e^{-\alpha i}$ (after normalization);
        \STATE Merge information subchannels ${\cal I}_{merge} = {\cal I}_{p1} \cup {\cal I}_{p2}$;
        \STATE Include additional subchannels ${\cal I}_{mutate}$ by sampling the remaining ones with probability $\beta$;
        \STATE Select $K$ information subchannels from ${\cal I}_{merge} \cup {\cal I}_{mutate}$ to generate an offspring ${\cal I}_{o}$;
        \STATE Evaluate ${\cal I}_{o}$ at $SNR_{eval}$ and insert back to population while maintaining ordering.
    \ENDWHILE
    \RETURN ${\cal I}_1, {\cal I}_2, \cdots, {\cal I}_M$
\STATE {\bf end function}
\end{algorithmic}
\caption{Multi-stage genetic algorithm for polar code construction}
\label{alg:genetic-multistage}
\end{algorithm}

\subsection{Pretraining example production}\label{section:int_learning:examples}
In terms of MDP, the code constructions generated by genetic algorithm represent good states with large reward values.
The remaining problem is how to design state-action $(s,a)$ pairs from these constructions.
One characteristic of this MDP is that, the state is a collection (set) of history actions taken to reach it.
Meanwhile, the order of actions is neither distinguishable from the state, nor relevant to reach the state.
Therefore, the state-action $(s,a)$ pairs can be produced based on two intuitions,
\begin{enumerate}
  \item Given a current state, if the agent can take one action to reach a good state, then this can be a valid state-action pair;
  \item If the current state is a good state, a potentially good choice of action can be the ones that has not been taken to reach the current state, while is recorded by some other states with close information length.
\end{enumerate}

The process to produce state-action $(s,a)$ pairs from good states is described in Alg.~\ref{Alg:pretrain}.
\begin{algorithm}
\caption{Pretraining example generation}
\label{Alg:pretrain}
\begin{algorithmic}[1]
\STATE {\bf function} example\_generation($polar\_codes\_constr$)
\STATE $examples=[]$
\STATE \# Based on intuition-1):
\FOR{$state \in polar\_codes\_constr$}
    \FOR{$n \in [0,N-1]$}
        \IF{$state(n)==1$}
            \STATE $data=state$
            \STATE $data(n)=0$
            \STATE $label=n$
            \STATE $examples.append([data,label])$
        \ENDIF
    \ENDFOR
\ENDFOR
\STATE \# Based on intuition-2):
\FOR{$state_s, state_d \in polar\_codes\_constr$ \AND $\left(\sum{state_d}==(1+\sum{state_s})\right)$}
    \FOR{$n \in [0,N-1]$}
        \IF{$\left(state_d(n)==1\right) \&\& \left(state_s(n)==0\right)$}
            \STATE $data=state_s$
            \STATE $label=n$
            \STATE $examples.append([data,label])$
        \ENDIF
    \ENDFOR
\ENDFOR
\RETURN $examples$
\STATE {\bf end function}
\end{algorithmic}
\end{algorithm}

\section{Evaluation}\label{section:evaluation}
In this section, we elaborate the model of the reinforcement learning algorithms, and evaluate the sample efficiency of various schemes.
\subsection{Model}\label{section:evaluation:model}
\begin{figure}[h]
\centering
    \includegraphics[width = 0.45\textwidth]{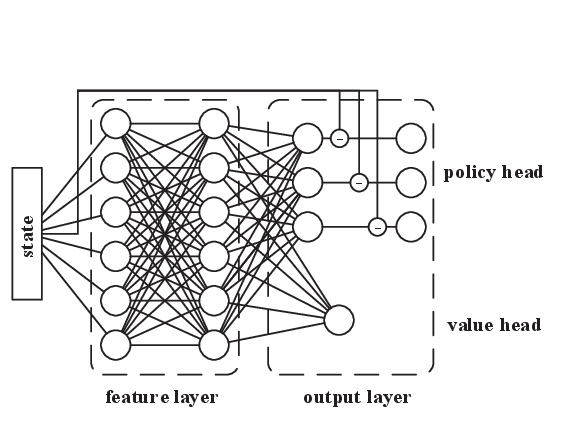}
    \caption{Neural network architecture.}
    \label{fig:NN_architecture}
\end{figure}

For the reinforcement learning algorithms, we use neural networks to represent the policy and value function.
The same neural network architecture, shown in Fig. \ref{fig:NN_architecture}, is deployed for all of the nested polar construction tasks.
For an input state $s$, a feed-forward network was used for feature extraction, with two fulled connected layers, with $2N$ tanh units per layer.
This feature layer was shared by policy and value function.
For the output layer, the policy function used a linear layer to screen out previously selected subchannels (e.g., by subtracting a larger value from the corresponding entries), and followed by a softmax nonlinearity to generate probability mass function (PMF).
The value function used a linear layer to output an estimated value for state $s$.

One synthesized loss function is used for simultaneous training of policy and value function,
\begin{equation}\label{Equ.Loss}
  Loss=Loss_A + \beta_c Loss_C - \beta_e H_A
\end{equation}
where $Loss_A$ and $Loss_C$ are the loss functions for actor and critic, and $H_A$ is the entropy for policy function, defined in section \ref{section:learning:RL_algs}. The policy function entropy is subtracted in loss function to encourage exploitation.

We define an action model and a training model, based on the agent's policy and value function.
The action model generates an action and a value estimation for an observed state.
For any legal state, the policy function calculates the probability mass function (PMF), based on which an action is randomly sampled.
The value function estimates the value for the state.
The training model trains the policy and value function.

\subsection{Speed optimizations}
Observing that MC simulations are time-consuming, we propose several optimizations to further improve sample efficiency.
\subsubsection{early termination}\label{section:learning:MDP:surrogate_env}
A ``\emph{surrogate environment}'' is defined to early terminate an episode at an ill-defined construction whose current reward is already bad.
At the beginning of the learning task, the entropy of agent's policy is large, therefore explorations are mostly random.
From the agent's perspective, when a state $s_t$ is encountered with a reward $r_t=-\log_{10}BLER \approx 0 (BLER \approx 1)$, we observed that the future rewards $r_{t'} (t'>t)$ are likely to be close to 0 except for the ones with $t'$ approaching $N$.
Since the agent can barely improve the policy when the reward values approximate 0, we can define a surrogate environment.
From the environment's perspective, this surrogate environment behaves exactly like the aforementioned base environment, except that when a state with reward value $r<r_{etthr}$ is reached, it returns the reward value along with a flag indicating the termination of current episode.
This environment is named ``\emph{base environment with early termination}'' and is a default option unless otherwise stated.
\subsubsection{memoization}\label{section:learning:MDP:memoization}
By definition, a state-reward pair corresponds to the performance of a specific construction. Once explored, it remains unchanged in this MDP task.
The most frequently encountered state-reward pairs are memoized for future retrieval.
This is shown to effectively reduce the MC simulation burden.
In addition, evaluating a code construction with a larger reward value, i.e. small BLER, requires longer MC simulation time. This is because more code blocks are simulated to collect sufficient errors.
Therefore, memoization is employed to collect state-reward $(s,r)$ pairs if the reward value $r>r_{recthr}$, such that future MC simulation is skipped if the same state (i.e., code construction) has been evaluated before.
\subsubsection{vectorized environment}\label{section:learning:MDP:vec_env}
A vectorized environment is defined to improve MC simulation efficiency, which is a collection of $n_{env}$ parallel base environments.
To guarantee independency among the environments, their random number generator seeds are set to different values.
For the vectorized environment, action model generates $n_{env}$ actions and $n_{env}$ value estimations based on the observed $n_{env}$ states from each base environment.
The training model trains the policy and value function, based on a batch of state-action-return-value $(s,a,R,V)$ tuples.
The batch size $n_{batch} = n_{env} \cdot n_{step}$, where $n_{step}$ is the timestep number for empirical return value estimation.

\subsection{Reinforcement learning}\label{section:evaluation:RL}
We conduct a series of experiments under SCL-Genie decoding to investigate the following questions:
\begin{enumerate}
  \item Which reinforcement learning algorithm is most sample efficient?
  \item How to select hyper-parameters to trade off between sample efficiency and convergence performance?
\end{enumerate}

The default parameters are listed in Table \ref{Table.Params}.
\begin{table}
  \centering
  \caption{Default parameters setting}
\begin{tabular}{|c|c|}
  \hline
  Parameters                                & values    \\ \hline
  polar code length                         & $N=256$   \\ \hline
  decoder                                   & SCL-Genie, SCL-PM \\ \hline
  SC list size                              & $L=8$     \\ \hline
  reward                                    & $r=-\log_{10}(BLER)$  \\ \hline
  BLER simulation error event count         & 1000      \\ \hline
  early termination                         & enable    \\ \hline
  early termination threshold               & $r_{etthr}=0.05$  \\ \hline
  clipping ratio in policy loss             & $\epsilon=0.2$    \\ \hline
  critic loss weight                        & $\beta_c=0.5$     \\ \hline
  entropy weight                            & $\beta_e=0$       \\ \hline
  learning rate                             & $3\cdot10^{-4}$   \\ \hline
  batch size                                & $n_{batch}=64$    \\ \hline
  feature extraction network                & 1024,1024         \\ \hline
  discount factor                           & $\gamma=0.2$      \\ \hline
  GAE factor                                & $\lambda=0.95$    \\ \hline
\end{tabular}
  \label{Table.Params}
\end{table}

\begin{figure}[h]
\centering
    \includegraphics[width = 0.45\textwidth]{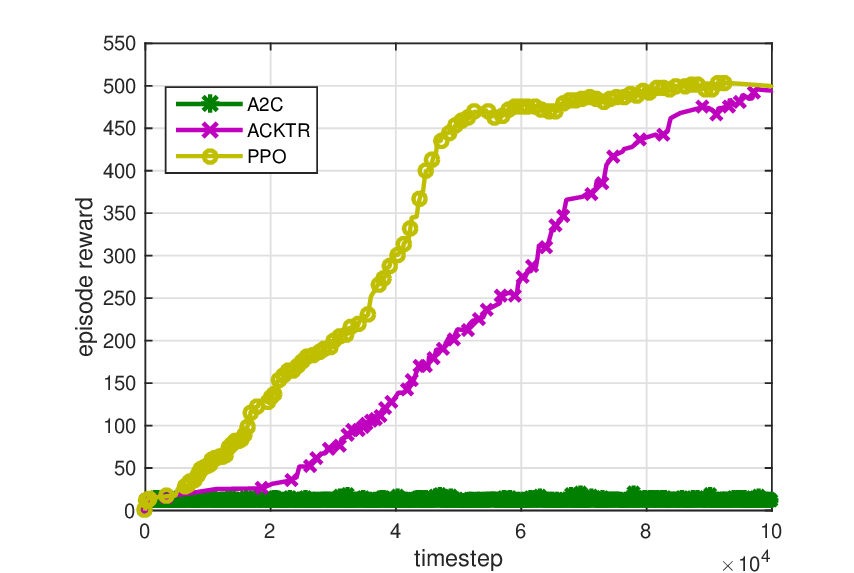}
    \caption{Sample efficiency comparison among A2C, ACKTR, and PPO. For each algorithm, episode rewards from 16 different base environments are plotted.}
    \label{fig:cmp_alg}
\end{figure}

Fig.~\ref{fig:cmp_alg} shows the episode rewards of A2C, ACKTR and PPO for 100E3 timesteps.
The number of timesteps for return estimation was optimized for each algorithm.
PPO outperformed A2C and ACKTR in terms of sample efficiency by a significant margin, and was therefore adopted in the following experiments.

\begin{figure}[h]
\centering
    \includegraphics[width = 0.45\textwidth]{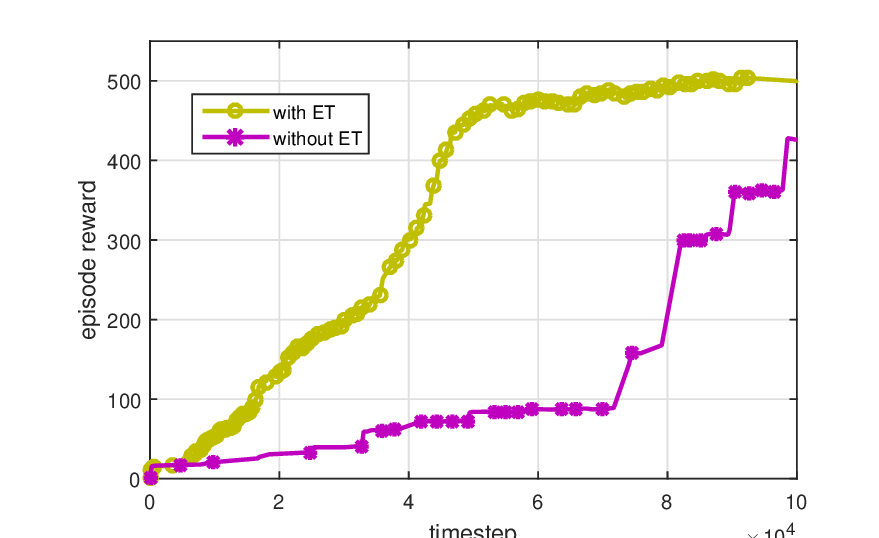}
    \caption{Sample efficiency comparison for base environment with and without early termination.}
    \label{fig:cmp_ET}
\end{figure}

Fig.~\ref{fig:cmp_ET} shows the episode rewards of base environment with and without early termination.
At the beginning of learning when episode reward is below $100$, base environment with early termination showed much better sample efficiency, since it saves the MC simulations of a number of trivial samples (with reward values approximating 0). Afterwards, the episode rewards for both base environments showed similar growing speed.
This proves that the early termination is effective and has little impact on the learning task except by skipping trivial sample simulations.

\begin{figure}[h]
\centering
    \includegraphics[width = 0.45\textwidth]{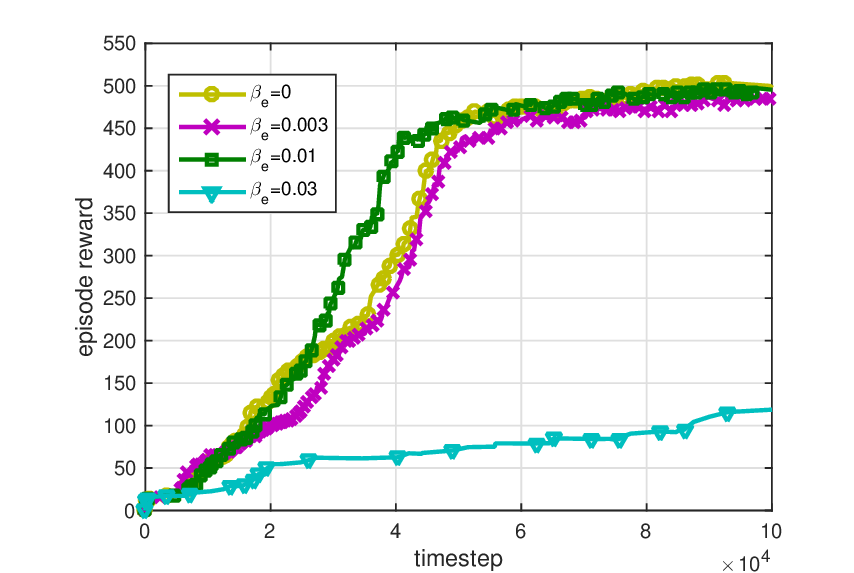}
    \caption{Sample efficiency comparison in terms of entropy weight.}
    \label{fig:cmp_Ent}
\end{figure}

Fig.~\ref{fig:cmp_Ent} shows the episode rewards for the amount of policy entropy evolved in loss function.
Similar sample efficiency is observed for entropy weight $\beta_e \leq 0.01$.
Nevertheless, entropy weight $\beta_e=0$ shows slightly better convergence performance.

\begin{figure}[h]
\centering
    \includegraphics[width = 0.45\textwidth]{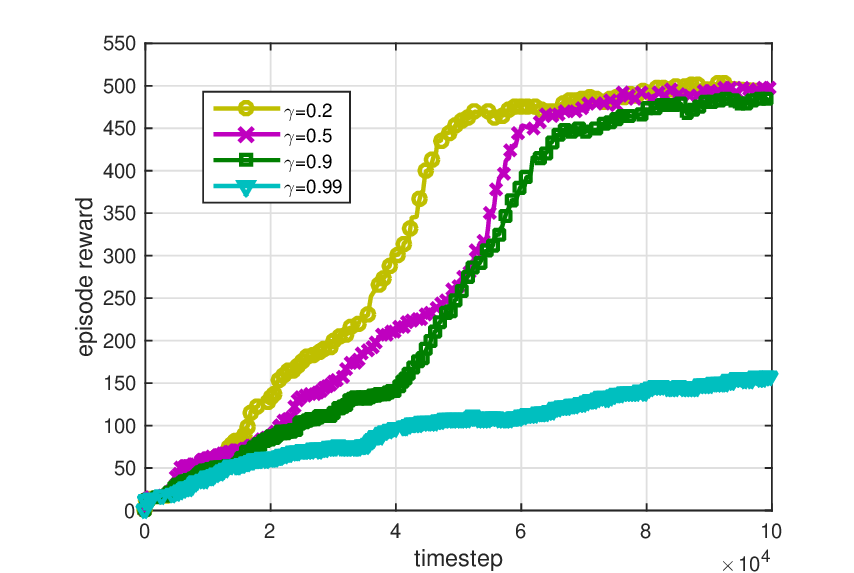}
    \caption{Sample efficiency comparison in terms of discount factor $\gamma$.}
    \label{fig:cmp_gamma}
\end{figure}

Fig.~\ref{fig:cmp_gamma} shows the episode rewards for discount factor selection.
For a smaller discount factor, the sample efficiency is increased since the current return would be affected by shorter future actions.
While for this learning task, the convergence performance is not compromised.

In this subsection, we demonstrate that PPO is sample efficient.
Early termination in the base environment saved MC simulations for trivial samples.
For entropy weight of $0$ and small discount factor, the sample efficiency is increased without compromising the learning performance.

\subsection{Integrated learning}\label{section:evaluation:pretrain}
In this subsection, we evaluate the integrated learning to show its improved sample efficiency.

We first obtained a population of polar codes for each $(N,K)$ pair by genetic algorithm.
Then we applied Alg.~\ref{Alg:pretrain} to generate examples of state-action $(s,a)$ pair.
The same policy network architecture is used in integrated learning as in Fig.~\ref{fig:NN_architecture}.
The policy network was trained on randomly sampled examples with stochastic gradient descent to minimize the training loss function,
\begin{equation}\label{Equ.Loss_pre}
  Loss_{pre}=Loss_{Apre} - \beta_{epre} H_{Apre}
\end{equation}
where $Loss_{Apre}$ is the cross entropy between policy output and the (one-hot) action label, $H_{Apre}$ is the entropy value of policy function, with entropy weight $\beta_{epre}=1.0$.
After 20 epoches of training, the policy function is saved for reinforcement learning as described in section~\ref{section:learning:RL_algs}.

\begin{figure}[h]
\centering
    \includegraphics[width = 0.45\textwidth]{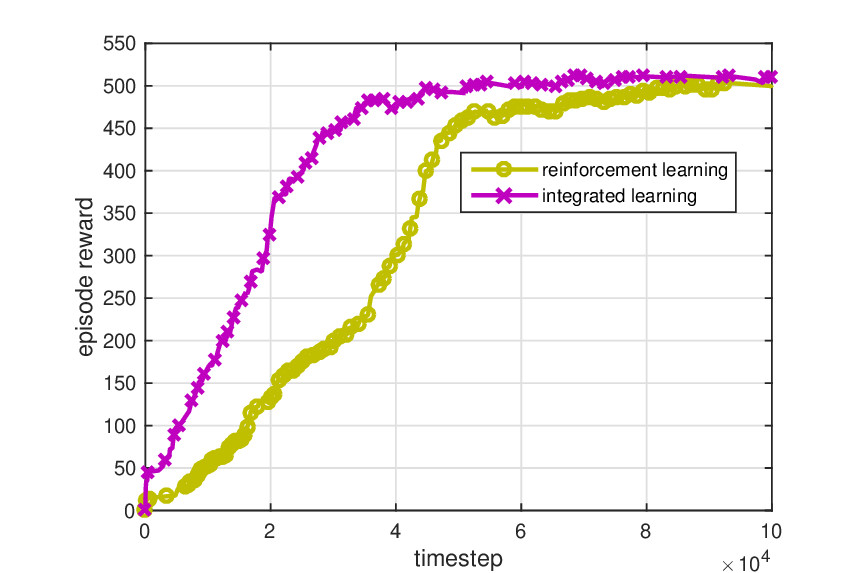}
    \caption{Sample efficiency comparison between reinforcement learning and integrated learning.}
    \label{fig:cmp_pretrain}
\end{figure}

Fig.~\ref{fig:cmp_pretrain} shows the episode rewards for 100E3 timesteps for reinforcement learning and integrated learning.
It is shown that integrated learning has better sample efficiency as well as larger episode reward values.

\subsection{BLER performance}\label{section:results}
For nested polar code construction with code length of $256$, the error correction performance of the learned codes are compared with those constructed by DE/GA. It should be noted that the comparison is unfair with respect to description and implementation complexity, because the constructions by DE/GA are not necessarily nested.

We consider two MDP tasks with different decoders:
\begin{itemize}
  \item SCL-Genie decoding under AWGN channel
  \item SCL-PM decoding under AWGN channel
\end{itemize}
The same learning method (parameters) are used for both MDP tasks.

\begin{figure}[h]
\centering
    \includegraphics[width = 0.45\textwidth]{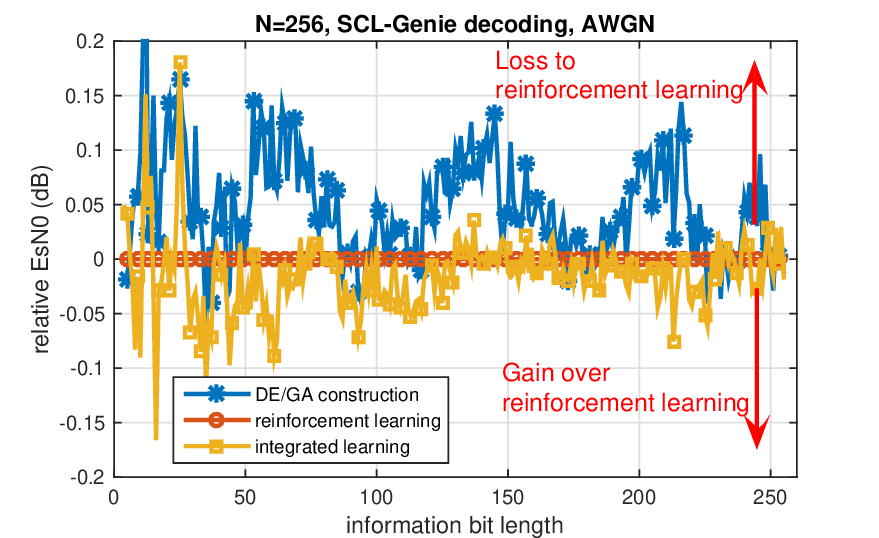}
    \caption{Relative performance between polar codes constructed by reinforcement learning and DE/GA for SCL-Genie decoding under AWGN channel.}
    \label{fig:cmp_SCLGenie}
\end{figure}

For SCL-Genie decoding under AWGN channel, nested polar code constructions are learned by reinforcement learning and integrated learning with 1E6 training timesteps.
Fig.~\ref{fig:cmp_SCLGenie} shows the relative EsN0 value (at BLER of $10^{-2}$) for the three constructions.
The nested polar constructions learned by both reinforcement learning and integrated learning outperform the case-by-case DE/GA constructions for a majority of cases.
Integrated learning exhibits even better performance, with a maximum gain over DE/GA approaching 0.3dB.

\begin{figure}[h]
\centering
    \includegraphics[width = 0.45\textwidth]{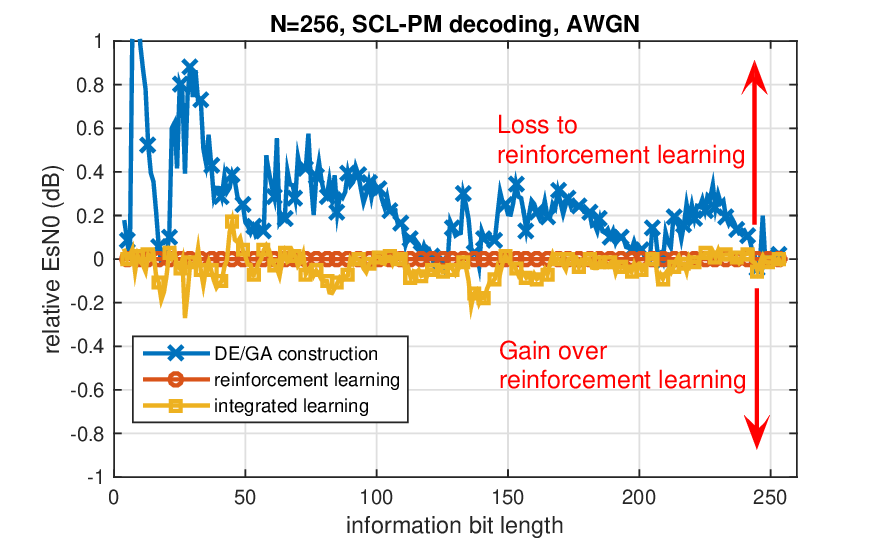}
    \caption{Relative performance between polar codes constructed by reinforcement learning and DE/GA for SCL-PM decoding under AWGN channel.}
    \label{fig:cmp_SCLPM}
\end{figure}

For SCL-PM decoding under AWGN channel, nested polar code constructions are learned with 100E3 training timesteps.
Fig.~\ref{fig:cmp_SCLPM} shows that the learned nested polar constructions outperform the case-by-case DE/GA constructions for almost all information length. The maximum performance gain achieves as large as 1.2dB.

\section{Conclusions}\label{section:conclusion}
In this paper, the ``constructor-evaluator'' framework in \cite{AI:AI_coding} is adopted to construct polar codes.
In particular, we show that constructing nested polar codes can be viewed as a Markov decision process.
Thus, reinforcement learning techniques such as A2C and its latest improvements (e.g., PPO and ACKTR) are employed to iteratively optimize the ``code construction'' policy without expert intervention.
An MDP environment is set up using the BLER performance as feedback to guide the learning process, which is implemented by neural network based policy and value functions.
To facilitate faster and better convergence, a multi-stage genetic algorithm is integrated in the RL algorithms to provide prior knowledge about (sub-)optimal code constructions.
We carry out extensive experiments to compare the learning process under various settings.
The polar code constructions for both SCL-PM and SCL-Genie decoders are obtained, which exhibit superior performance over classic constructions.

\ifCLASSOPTIONcaptionsoff
  \newpage
\fi


\begin{thebibliography}{1}
\bibitem{Shannon}
C. E. Shannon, ``A mathematical theory of communication'', \textit{Bell System Technical Journal}, vol. 27, no.34, pp. 379--423, Jul. 1948.
\bibitem{AI:AI_coding}
L. Huang, H. Zhang, R. Li, Y. Ge, J. Wang , ``AI Coding: Learning to Construct Error Correction Codes'', \textit{arXiv:1901.05719 }, Jan. 2019.
\bibitem{AI:RL_LDPC}
M. Zhang, Q. Huang, S. Wang, Z. Wang, ``Construction of LDPC Codes Based on Deep Reinforcement Learning'', \textit{ 2018 10th International Conference on Wireless Communications and Signal Processing}, Oct 2018.
\bibitem{AI:GeneAlg_Polar}
A. Elkelesh, M. Ebada, S. Cammerer, S. Brink , ``Decoder-tailored Polar Code Design Using the Genetic Algorithm'', \textit{arXiv:1901.10464 }, Jan. 2019.
\bibitem{AI:GeneAlg_LDPC}
A. Elkelesh, M. Ebada, S. Cammerer, S. Brink , ``Decoder-in-the-Loop: Genetic Optimization-based LDPC Code Design'', \textit{arXiv:1903.03128 }, Mar. 2019.
\bibitem{Polar:nested}
S. Korada and R. Urbanke, ``Polar codes are optimal for lossy source coding'', \textit{Transactions on Information Theory}, vol. 56, no. 4, pp. 1751-1768, Mar. 2010.
\bibitem{Polar:Arikan}
E. Arikan, ``Channel polarization: A method for constructing capacity-achieving codes for symmetric binary-input memoryless channels'', \textit{IEEE Trans. Inf. Theory}, vol. 55, no. 7, pp. 3051--3073, Jul. 2009.
\bibitem{Polar:DE1_Mori}
R. Mori and T. Tanaka, ''Performance of polar codes with the construction using density evolution'', \textit{IEEE Communications Letters}, vol. 13, no. 7, pp. 519--521, July 2009.
\bibitem{Polar:DE2_Mori}
R. Mori and T. Tanaka, ''Performance and construction of polar codes on symmetric binary-input memoryless channels'', \textit{IEEE International Symposium on Information Theory}, June 2009.
\bibitem{Polar:DE3_Tal}
I. Tal and A. Vardy, ``How to construct polar codes'', \textit{IEEE Trans. Inf. Theory}, vol. 59, no. 10, pp. 6562--6582, July 2013.
\bibitem{Polar:DE4_Pedarsani}
R. Pedarsani, S. Hassani, I. Tal and E. Telatar, ``On the construction of polar codes'', \textit{IEEE International Symposium on Information Theory}, July 2011.
\bibitem{Polar:GA_Trifonov}
P. Trifonov, ``Efficient design and decoding of polar codes'', \textit{IEEE Transactions on Communications} vol. 60, no. 11, pp. 3221--3227, Nov. 2012.
\bibitem{Polar:SCL}
I. Tal and A. Vardy, ``List decoding of polar codes'', \textit{IEEE International Symposium on Information Theory Proceedings} pp. 1--5, 2011.
\bibitem{Polar:SCS}
K. Chen, K. Niu and J. Lin, ``Improved Successive Cancellation Decoding of Polar Codes'', \textit{IEEE Transactions on Communications} vol. 61, no. 8, pp. 3100-3107, August 2013.
\bibitem{Polar:212}
3GPP, ``NR; Multiplexing and channel coding'', \textit{3GPP TS 38.212}, 15.5.0, Mar. 2019.
\bibitem{RL:PPO}
J. Schulman, F. Wolski, P. Dhariwal, A. Radford, O. Klimov, ``Proximal Policy Optimization Algorithms'', \textit{arXiv:1707.06347 }, Jul. 2017.
\bibitem{RL:ACKTR}
Y. Wu, E. Mansimov, S. Liao, R. Grosse, J. Ba, ``Scalable trust-region method for deep reinforcement learning using Kronecker-factored approximation'', \textit{arXiv:1708.05144 }, Aug. 2017.
\bibitem{RL:TRPO}
J. Schulman, S. Levine, P. Moritz, M. Jordan, P. Abbeel, ``Trust Region Policy Optimization'', \textit{arXiv:1502.05477 }, Feb. 2015.
\bibitem{RL:GAE}
J. Schulman, P. Moritz, S. Levine, M. Jordan, P. Abbeel, ``High-Dimensional Continuous Control Using Generalized Advantage Estimation'', \textit{arXiv:1506.02438 }, Jun. 2015.
\bibitem{RL:Pretrain1}
D. Silver, A. Huang, C. J. Maddison, et al., ``Mastering the game of Go with deep neural networks and tree search'', \textit{Nature} vol. 529, pp. 484, 2016.
\bibitem{RL:Pretrain2}
X. Zhang, H. Ma, ``Pretraining Deep Actor-Critic Reinforcement Learning Algorithms With Expert Demonstrations'', \textit{arXiv:1801.10459 }, Jan. 2018.
\bibitem{RL:Pretrain3}
Y. Gao, H. Xu, J. Lin, F. Yu, S. Levine, T. Darrell, ``Reinforcement Learning from Imperfect Demonstrations'', \textit{arXiv:1802.05313 }, Feb. 2018.
\end{thebibliography}
\end{document}